# Bosonic condensation of exciton-polaritons in an atomically thin crystal

Carlos Anton-Solanas[1,2,*,†], Maximilian Waldherr[1,*], Martin Klaas[1], Holger Suchomel[1], Hui Cai[3], Evgeny Sedov[4,5,6], Alexey V. Kavokin[4,5,7], Sefaattin Tongay[8], Kenji Watanabe[9], Takashi Taniguchi[10], Sven Höfling[1,11] and Christian Schneider[1,2,†]

[1]*Technische Physik and Würzburg-Dresden Cluster of Excellence ct.qmat, Universität Würzburg, D-97074 Würzburg, Am Hubland, Germany.*
[2]*Carl von Ossietzky University, 26129 Oldenburg, Germany.*
[3]*University of California, Merced, California, USA*
[4]*School of Science, Westlake University, 18 Shilongshan Road, Hangzhou 310024, Zhejiang Province, People's Republic of China*
[5]*Institute of Natural Sciences, Westlake Institute for Advanced Study, 18 Shilongshan Road, Hangzhou 310024, Zhejiang Province, People's Republic of China*
[6]*Vladimir State University named after A. G. and N. G. Stoletovs, Gorky str. 87, 600000, Vladimir, Russia*
[7]*Spin Optics Laboratory, St-Petersburg State University, Russia.*
[8]*School for Engineering of Matter, Transport, and Energy, Arizona State University, Tempe, Arizona 85287, USA*
[9]*Research Center for Functional Materials, National Institute for Materials Science, 1-1 Namiki, Tsukuba 305-0044, Japan*
[10]*International Center for Materials Nanoarchitectonics, National Institute for Materials Science, 1-1 Namiki, Tsukuba 305-0044, Japan*
[11]*SUPA, School of Physics and Astronomy, University of St. Andrews, St. Andrews UK.*

*These authors contributed equally to this work

†Corresponding author. Email: carlos.anton-solanas@uni-oldenburg.de, christian.schneider@uni-oldenburg.de

**The emergence of two-dimensional crystals has revolutionized modern solid-state physics. From a fundamental point of view, the enhancement of charge carrier correlations has sparked enormous research activities in the transport- and quantum optics communities. One of the most intriguing effects, in this regard, is the bosonic condensation and spontaneous coherence of many-particle complexes. Here, we find compelling evidence of bosonic condensation of exciton-polaritons emerging from an atomically thin crystal of $MoSe_2$ embedded in a dielectric microcavity under optical pumping. The formation of the condensate manifests itself in a sudden increase of luminescence intensity in a threshold-like manner, and a significant spin-polarizability in an externally applied magnetic field. Spatial coherence is mapped out via highly resolved real-space interferometry, revealing a spatially extended condensate. Our device represents a decisive step towards the implementation of coherent light-sources based on atomically thin crystals, as well as non-linear, valleytronic coherent devices.**

Bose Einstein Condensation (BEC) represents a non-classical phase transition, characterized by the collapse of an ensemble of quantum particles into a macroscopic and coherent state. Experimentally, the investigation of BECs was pioneered in the field of ultra-cold atoms[1,2]. However, it has been proposed early[3], and subsequently verified that bosonic many-body excitations in solids[4], including excitons and exciton-polaritons, are capable of forming non-equilibrium condensates at relatively high temperatures which makes them a more user-friendly system for such fundamental studies. The solid-state, 'on-chip' character intrinsically yields an aspect of practicality, and the fact that excitons and exciton-polaritons emit light makes their condensates of high appeal in the application-driven research for novel and innovative light sources[5], as well as in the field of non-linear photonics[6]. The latter is particularly relevant for the case of exciton-polaritons formed in high-quality microcavities loaded with active materials in the strong light-matter coupling regime[7].

The non-linear character, as well as the spinor properties which dominate the condensation behaviour of exciton-polaritons and affect their intrinsic properties, are strongly linked to the materials embedded in the microcavities. Recently, atomically thin crystals of transition metal dichalcogenides (TMDCs) have emerged as a new, compelling platform in solid-state-based cavity quantum electrodynamics[8], benefitting from ultra-stable excitons[9], giant oscillator strength, and exotic polarization and topological properties.

A variety of reports have now addressed the emergence of exciton-polaritons with single, and multiple TMDC crystals[10–13] embedded in high-quality factor microcavities. Later, aspects in association with the valley-character of excitons in TMDC layers in the strong coupling regime has been demonstrated[14,15]. However, despite recent discoveries of strong non-linearities of TMDC excitons[16–18], phenomena related to bosonic condensation of exciton-polaritons with TMDC crystals, thus far, remain unaddressed. Hints to bosonic condensates of bare excitons in TMDC van-der-Waals heterostructures were recently reported[19], however the strong inhomogeneities of the utilized samples prohibited the exploration of spatially extended coherent states.

Here, we report the emergence of a coherent condensate of exciton-polaritons in a microcavity loaded with a single crystal of monolayer $MoSe_2$. Our device, which is operated at cryogenic temperatures (4 K) features the sharp non-linear threshold characteristic for a polariton laser and demonstrates a strong valley polarization in an externally applied magnetic field in the regime of bosonic condensation. Finally, we observe distinct features of spatial coherence via the classical interferometry.

**Sample and experimental setup**

The studied sample structure is schematically depicted in Fig. 1a. It is based on a hybrid III/V-dielectric cavity, which was assembled mechanically. The bottom distributed Bragg reflector (DBR), which was grown by molecular beam epitaxy, consists of 24 $AlAs/Al_{0.2}Ga_{0.8}As$ mirror pairs with thicknesses of 62 and 53 nm, respectively, and features a stop band centered around 753 nm. A 52 nm thick AlAs spacer forms the lower half of the optical cavity. The spacer is capped by a 3 nm GaAs capping layer and features a heavily doped GaAs quantum well (4.75 nm thick) located 10 nm underneath the surface. The inclusion of GaAs layers in the heterostructures significantly improves the crystal quality at the GaAs/TMDC interface, and provides access to carriers to enhance polariton-electron scattering processes.

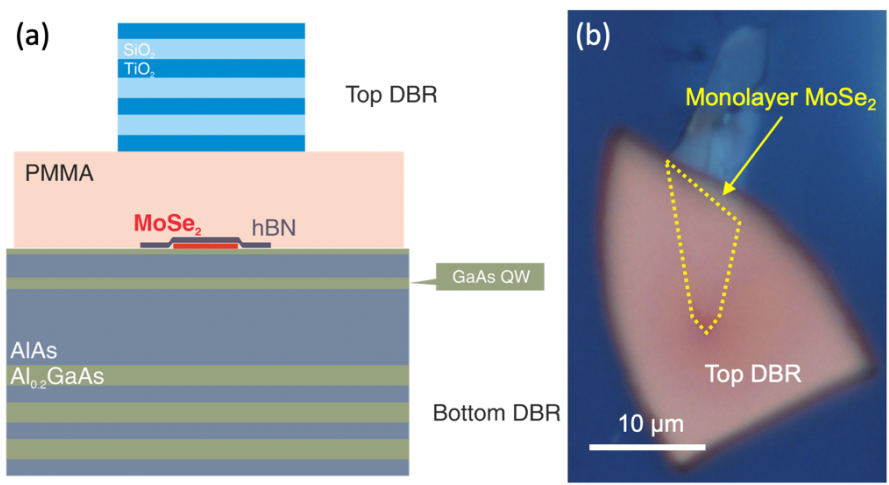

Fig. 1. **Sample structure and map**. (a) Schematic drawing of the microcavity structure. The lower DBR was grown epitaxially. The $MoSe_2$ layer is covered by a thin layer of hBN and a layer of PMMA that acts

as a spacer. The top DBR is composed of 9.5 pairs of $TiO_2/SiO_2$ and was mechanically separated from a carrier substrate and transferred on top of the PMMA (see details in the text). (b) Microscope image of the full structure. The position of the monolayer underneath the top DBR is schematically indicated by the dotted yellow line.

Next, an atomically thin layer of a $MoSe_2$ isolated from a CVD-grown bulk crystal is transferred onto the bottom DBR via the dry-gel stamping method[20]. Subsequently, it is capped by a thin layer of hexagonal boron nitride (hBN). Then, we spin-coated an ~90 nm thick poly-methyl-methacrylate (PMMA) buffer layer, forming the upper half of the cavity. Finally, a piece of a separate $SiO_2/TiO_2$ DBR with lateral dimensions of a few tens of μm (composed by 8.5 pairs, stop band center at 750 nm, terminated by $TiO_2$ on both sides) is mechanically peeled off its substrate and transferred onto the buffer layer with the same dry-gel method, concluding the heterostructure sketched in Fig. 1(a). Figure 1(b) shows a microscope image of the final structure (top view), where the contour of the $MoSe_2$ monolayer (4 μm x 12 μm) is marked by the yellow dotted line.

### Dispersion relation and detuning characteristics

We study our sample at cryogenic temperatures via momentum-resolved micro-PL-spectroscopy[21] in the back-Fourier plane imaging configuration. The lower part of Fig. 2(a) depicts the dispersion relation of the lower polariton branch, resulting from the coupled TMDC monolayer-microcavity resonance. The energy of the picosecond pulsed laser was chosen as 1.6714 eV to achieve a compromise between signal strength and accessibility of the energy range. The excitation power is set well below threshold.

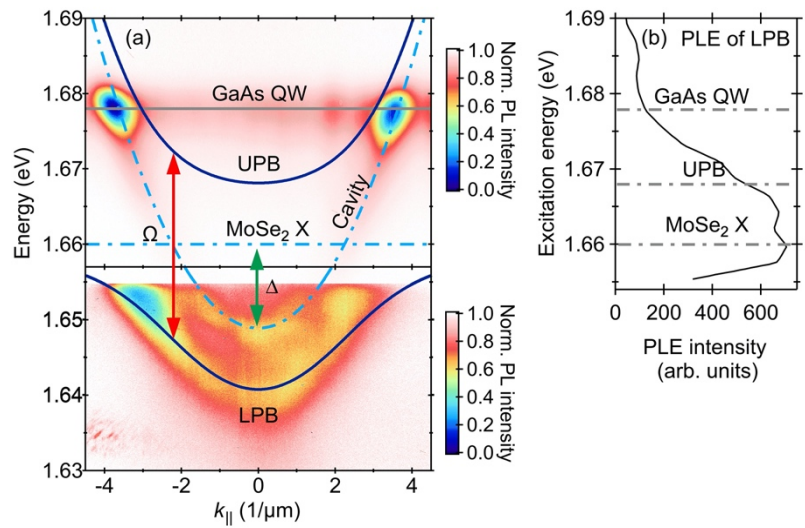

Fig. 2. **Optical properties of TMDC polaritons**. (a) Dispersion relation of the sample, showing the lower (LP) and upper (UP) polariton branches, $MoSe_2$ exciton and cavity modes and the GaAs QW. The upper part (from 1.657 eV) of the figure is recorded under continuous wave excitation with an energy of 1.705 eV. For the lower part (below 1.657 eV), the excitation is changed to picosecond pulses tuned at 1.6714 eV. For the sake of clarity, the intensity scale is normalized separately for both images. (b) PL emission intensity of the lower polariton branch as a function of the energy of the continuous wave excitation laser. Remarkably, the emission only increases for excitation energies below the QW resonance, demonstrating that there is no significant coupling between the QW and the polariton system.

The PL map features two curved dispersion relations. However, only the low energy branch can be detected on the position of the monolayer, whereas the higher energy luminescence (1.65 eV at $k_{||} = 0$) persists at sample positions in the periphery of the monolayer crystal (see

Fig. S1 in the Supp. Material). Thus, we conclude, that the high-energy branch arises from luminescence of the uncoupled photonic resonance defined by the optical microcavity. From a reference sample without the top DBR (see Fig. S2 of the Supplementary Material), we can accurately determine the spectral position of the absorption of the MoSe$_2$ monolayer/hBN stack (1.660 eV). We fit the lower polariton branch (termed LPB in Fig. 2a) in the framework of a standard coupled oscillator approach (see supplementary section S1). The model yields a normal-mode-splitting of 25±2 meV, which is in general agreement with earlier reports on strongly coupled MoSe$_2$–cavity structures[22–24]. Indeed, alongside the profound absorption feature from the MoSe$_2$ exciton, we find indications of absorption from the upper polariton branch in a photoluminescence excitation (PLE) measurement depicted in Fig. 2(b). We note that no significant absorption is detected from the doped GaAs QW to the polaritonic state (see horizontal dot-dashed lines in Fig. 2b) indicating the resonance energies of the QW, the upper polariton and the MoSe$_2$ exciton). Further, conducting a momentum resolved photoluminescence experiment with 6 mW of continuous wave excitation at 1.705 eV, see top part of Fig. 2a), we clearly observe the crossing (and no coupling) of the QW-luminescence peak with the empty cavity dispersion branch.

This demonstrates that the GaAs QW does not contribute to the strong coupling and, in contrast to earlier works discussing hybrid TMD-QW polaritons[19], this polaritonic state solely emerges from the strong coupling conditions between the MoSe$_2$ monolayer and the cavity mode.

**Condensation threshold**

The high-density behaviour of our device is captured in Fig. 3, where we study its nonlinear input-output characteristic. In this pump power dependence measurement, we excite with linearly-polarised picosecond laser pulses at a repetition rate of 80 MHz. The laser is tuned to an energy of 1671.4 meV (~40 meV above the polariton emission) to overlap with the upper polaritonic mode. In panels (a-c), we show the dispersion relation maps before (a) and after condensation threshold (b,c); the relative false colour scale of the maps reveals a strong nonlinear increase of the emission at the threshold power (14 mW).

The full pump power dependence is represented in panel (d), where the photoluminescence of the dispersion relation has been integrated for each power, providing the characteristic representation of the polariton non-linear threshold. We note that both vertical and horizontal axes are in logarithmic scale and the emission intensity increase between the two pump powers right below and above threshold is by more than a factor of 4.

In Fig. 3(e) we plot the extracted peak energy as a function of the pump power. At threshold, we observe a significant jump of the peak position by -5 meV. This strongly suggests, that our condensate is formed in a localized state of a lower energy, which is induced by disorder or inhomogeneous strain in our system. Such behaviour is well-understood and indeed commonly observed in disordered microcavities[25], structured cavities[26,27] and devices comprising strain traps[28]. Furthermore, the lower polariton mode displays a blueshift above threshold, which amounts to ~2 meV as determined by analysing the statistical maximum of the luminescence peak. However, we note that this mode has a distinct sub-structure, which most likely arises from spatial sample inhomogeneities. As we show in the supplementary section of the manuscript (Fig. S3), this substructure of the peak results in an over-estimated blueshift, such that the magnitude of 2 meV can only serve as an upper boundary.

The overall emission linewidth of the photoluminescence is depicted in Fig. 3f. Notably, we observe an increase in the emission linewidth towards the threshold, followed by a sudden drop at threshold to a value of ~12 meV. Indeed, this value remains constant with increasing pump power, which indicates that it is intrinsically limited by sample disorder and does not reflect the temporal coherence properties of the device emission. This is further evidenced by a large deviation of the spectral shape of the emission from a Lorentzian shape above the threshold.

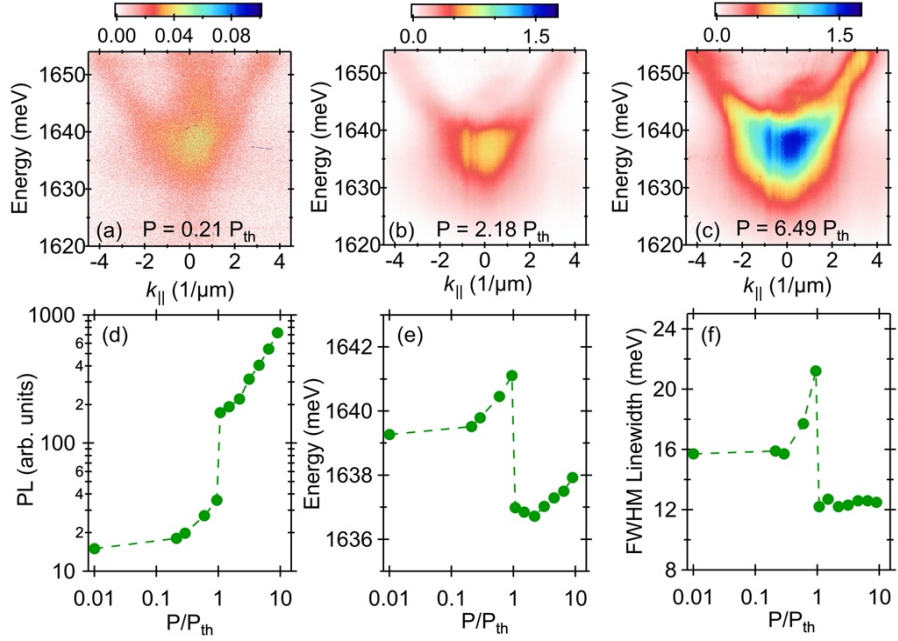

Fig. 3. **Nonlinear polariton emission, emission intensity versus pump power**. (a-c) Polariton dispersion relation maps encoded in a false colour scale for pump powers of 0.21, 2.18 and 6.49 $P_{th}$, respectively. (d-f) Integrated photoluminescence (logarithmic scale), energy of emission and linewidth as a function of pump power (logarithmic scale), respectively.

## Polariton coherence

One of the most important properties in a macroscopic condensed state is the emergence of spatially-extended phase coherence. The degree of spatial coherence can be investigated via self-interfering the polariton emission in a Michelson interferometer. In such an interferometer, the relative temporal delay between the arms can be tuned within the time-scale of the polariton emission lifetime, and where one of the arms contains a retro-reflector, which inverts both spatial axes of the reflected image.

Provided the cavity quality factor (~500, as a lower bound) and the emission characteristics of our TMD excitons (with a lifetime of ~390 fs[29]), we expect a sub-picosecond polariton lifetime of ~210 fs. The zero-delay calibration of the interferometer is initially set using the pulsed laser. This allows us to retrieve the corresponding zero delay of the polariton emission. Initially, we record the real space interferograms of the polariton emission above and below threshold (see Fig. 4(a,b), corresponding to 9 $P_{th}$ and 0.2 $P_{th}$, respectively) at zero delay, with retro-reflected images and displaced by 2 microns. An extended interference pattern appears in when above threshold (Fig. 4a), being several times larger than the pump spot (which has a FWHM of ~2 µm). On the contrary, in below threshold (Fig. 4b) we barely observe interference fringes since no macroscopic coherence is formed.

The Fourier-transform analysis of such interferograms, combined with the intensity maps of each interferometer arm, allows to reconstruct the real space distribution of the first order coherence, also known as the $g^{(1)}(\Delta t)$ function. This analysis is shown in Figs. 4(c-e), where we present different $g^{(1)}$ distributions as a function of three different temporal delays (0, 0.15 and 0.42 ps, respectively). The pump power is kept at 9 $P_{th}$. The $g^{(1)}$ map shown in panel (c) corresponds to the interferogram displayed in panel (a).

We systematically measure the polariton $g^{(1)}$ maps versus delay in a temporal window of 1 ps around zero delay. From these maps, we extract the averaged $g^{(1)}$ value in a 1 µm diameter circle around the origin of coordinates (indicated by the white circle in panel (e)) and plot it versus the delay, see Fig. 4(f). We observe a temporal Gaussian profile with a coherence time

(FWHM) of 360±10 fs, which, as expected for the case of pulsed excitation, is in the same order of magnitude as the previously-indicated polariton lifetime of the system [30,31].

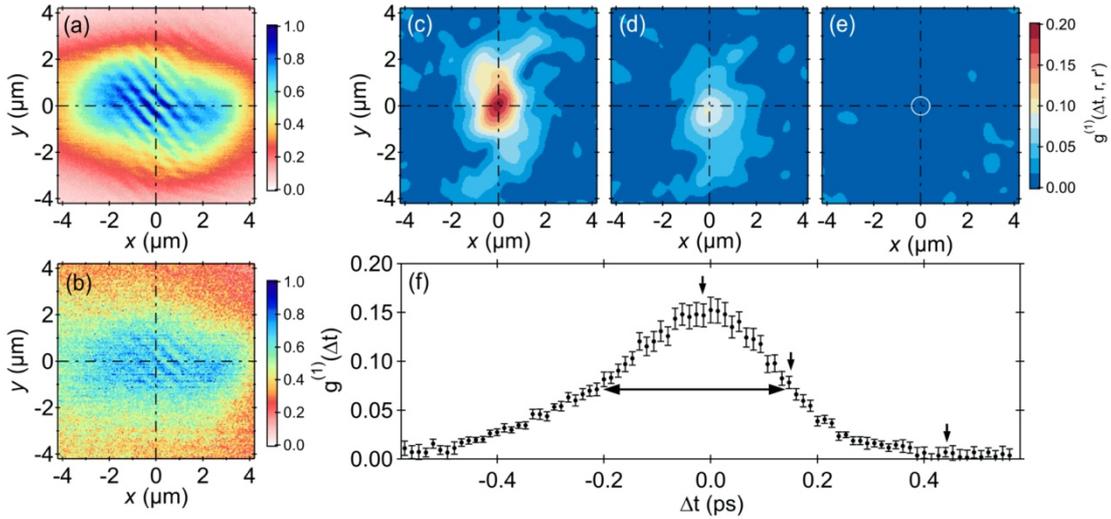

Fig. 4. **First order autocorrelation measurement of the polariton condensate**. (a,b) Real space distribution of the polariton emission interference at zero delay at 9$P_{th}$ and 0.2$P_{th}$, respectively. (c-e) Real space distribution of the first order correlation function ($g^{(1)}$) for three different delays: zero, 0.15 and 0.42 ps (same excitation conditions as panel (a)). (f) $g^{(1)}$ versus delay between the interferometric arms, the value of $g^{(1)}$ is the average value within the white circle depicted in panel (e), the error bars correspond to the mean standard deviation. Vertical black arrows highlight the delays of the $g^{(1)}$ maps represented in panels (c-e), the horizontal arrow indicates the temporal coherence length at half maximum of the $g^{(1)}$ function.

**Valley Control of TMDC polaritons**

In MoSe$_2$ monolayers, the K and K' valleys with opposite spins are non-degenerate due to the lack of an inversion centre. This causes the spin of an exciton to be intrinsically locked to the corresponding valley, opening up the possibility to manipulate the valley degree of freedom by means of optical injection, as well as built-in and external magnetic fields. Valley polarization, thus, manifests itself in the increase of the degree of circular polarization of the device emission. In Fig. 5 we study the response of our polaritonic condensate in the presence of a vertical external magnetic field. By injecting the condensate (~8.5 $P_{th}$) with a linearly polarized laser, we ensure a balance between K and K' polaritons at zero magnetic field.

Figures 5(a-c) depict dispersion relations recorded at -9, 0, and 9 T for a constant pump power of 120 mW. Despite the observation of an increase of the condensation threshold with applied magnetic field, similar to previous reports of GaAs polariton condensates[32], our pump power was sufficiently strong to remain in the non-linear regime of bosonic condensation throughout the entire experimental series. In Figs. 5(d-f), we plot the degree of circular polarisation ($S_z$) for the corresponding magnetic fields. The maps of $S_z$ distribution along the low polariton branch display a significant change from negative to positive values while the magnetic field varies from -9 to 9 T. In the extreme cases (panels d and f), both $S_z$ maps show an inhomogeneous distribution arising from the wide linewidth of the polariton emission (~12 meV), with predominantly higher $S_z$ values for lower energies, where the energy splitting between σ$^+$ and σ$^-$ populations becomes more evident.

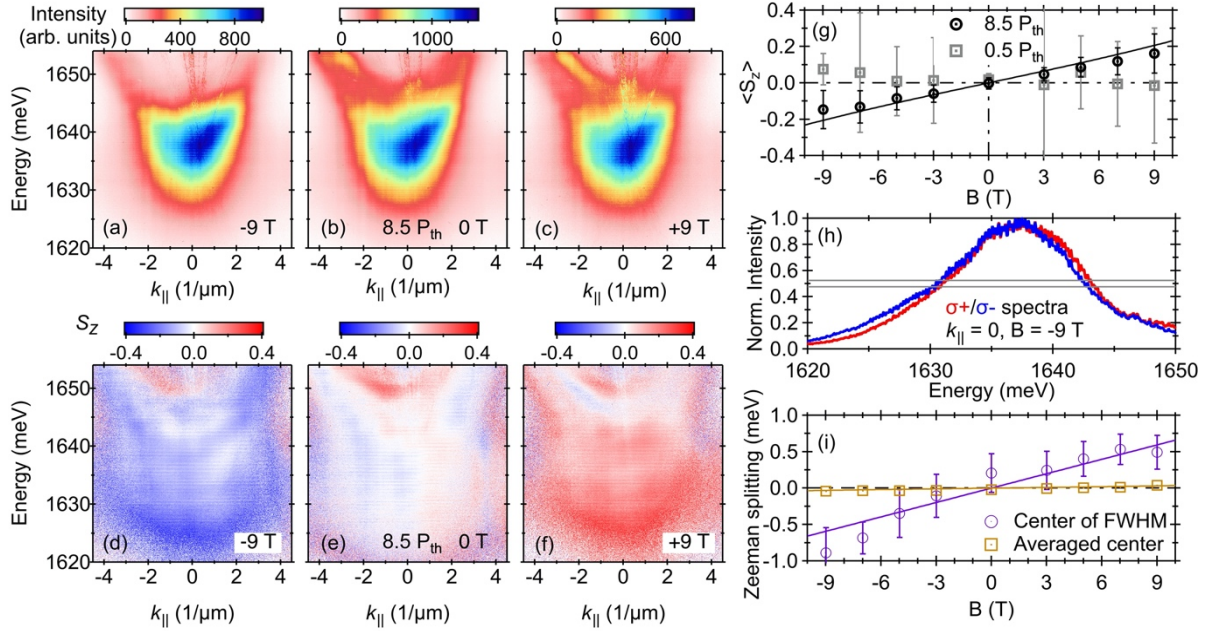

Fig. 5. **Dispersion relation and circular degree of polarisation under strong magnetic fields**. (a-c) PL intensity maps, $\sigma^+ + \sigma^-$, and (d-f) degree of circular polarization, $S_z$, as a function of energy and momentum ($k_\parallel$) of the polariton condensate under a pump power of 8.5 $P_{th}$, encoded in two different colour scales, respectively. (a,d), (b,e) and (c,f) panels correspond to magnetic fields of -9, 0 and 9 T. (g) Spectrally-averaged value of $S_z$ at $k_\parallel = 0$ (black circles) as a function of magnetic field under a pump power of 8.5 $P_{th}$. The corresponding values for $S_z$ at a pump power of 0.5 $P_{th}$ are shown as grey squares (see Fig. S4 of the Supp. Material). (h) Normalised $\sigma^+/\sigma^-$ spectra in red/blue at $k_\parallel = 0$ and a magnetic field of -9 T. The horizontal grey lines at half maximum highlight the spectral shift between $\sigma^+/\sigma^-$ detections. The energy shift between $\sigma^+/\sigma^-$ FWHMs is used to extract the Zeeman splitting represented in purple open circles in Fig. 5(h). (g) Extracted Zeeman splitting under two different criteria: (purple circles) energy splitting of the normalised $\sigma^+/\sigma^-$ spectra at full-width half maximum and (yellow squares) energy splitting extracted from the averaged centre of the intensity of $\sigma^+/\sigma^-$ spectra at $k_\parallel = 0$ (see Fig. S5 of the Supp. Material for further details). The full lines in panels (g,i) are theoretical predictions from the model described in the supplementary section S2.

A systematic variation of the external magnetic field displays progressive valley polarization of the polaritonic condensate in Fig. 5(g). Averaging $S_z$ at $k_\parallel = 0$, in the range of energies 1620-1650 meV, we retrieve a clear linear variation of $<S_z>$ as a function of magnetic field (black circles). For the sake of comparison, we include in the same panel the corresponding analysis for a pump power of 0.5 $P_{th}$ (grey squares). The error bars in both cases correspond to the standard deviation of the spectrally averaged value of $S_z$. Interestingly, the below-threshold behaviour does not show a strong magnetic field dependence, having a greater error than the $<S_z>$ value itself. This hints at the acceleration of polariton scattering via final state stimulation above threshold conditions, which in part prevents the rapid valley depolarization in the linear regime.

Last, in Figs. 5(h,i) we investigate the energy splitting of the polariton resonance in the presence of the magnetic field. Due to the intrinsically broad spectrum of the polariton cloud and its complex shape, the extraction of the Zeeman splitting of the $\sigma^+/\sigma^-$ emission requires some careful analysis.

In first place, we compare the normalised $\sigma^+/\sigma^-$ spectra at $k_\parallel = 0$ at different magnetic fields (see Fig. 5(h) where we show the particular case of B = -9 T). While the maximum value of these $\sigma^+/\sigma^-$ spectra only displays a small shift in energy, we notice a clear lateral offset of the spectra in their intensity wings. We extract the energy splitting from the energy difference between centres of the FWHM of each σ⁺/σ⁻ spectra. The grey lines at 0.5 normalised intensity highlight the region where the blue and red curves are shifted in energy by ~1 meV. Applying the same analysis for all the measured magnetic field values, we retrieve the Zeeman splitting depicted in purple circles in Fig. 5(i), where the fitted slope (full purple line) is 0.078±0.008 meV/T.

The magnetic field-induced splitting can be related to the exciton $g$-factor via the expression $|X|^2 g\mu_B B$, where $g\mu_B B$ is the splitting strength of the reservoir excitons, $g$ denotes the excitonic $g$-factor, and $|X|^2$ the excitonic fraction of the lower polariton branch, which in our case is 30% at $k_\parallel = 0$ (given the negative detuning and Rabi splitting previously discussed in Fig. 1). From the slope in Fig. 5(i), we obtain an exciton g-factor of 4.5±0.5, which is in good agreement with previously reported values for MoSe$_2$ excitons[33].

We notice, that a peak analysis based on evaluating the statistical centre of each normalized normalised σ⁺/σ⁻ spectra yields a Zeeman splitting, yet with a slope ~20 times smaller (yellow squares Fig 5(i)). This method, being more punitive when applied to signals of complex shape (as in our case), serves as a lower bound for the estimation of the magnetic field splitting of our polariton condensate.

To support our experimental findings, we provide the simulation of the effect of the external magnetic field on the degree of circular polarization of polaritons. We use the model proposed in Ref. [34] which is based on the system of semiclassical rate equations for the occupations of the condensed and non-condensed polariton states taking into account their spin-polarization (see supplementary section S2 of the manuscript). The lines in Fig. 5(g,i), fitting the experimental data, are consistent with the theory, which confirms the build-up of the circular polarization of the polariton state in the presence of a magnetic field accounting for the spin relaxation and stimulated scattering processes.

In conclusion, we have demonstrated the bosonic condensation of exciton-polaritons formed from strong coupling between monolayer MoSe$_2$ excitons and a microcavity-photonic field. The input-output response of the system reveals a clear switching threshold, with a sudden increase of the non-linear emission, linewidth narrowing and a slight blueshift of emission energies above threshold, all of them highlighting the emergence of polariton lasing. The first order coherence function of the emission shows a macroscopically extended common phase, which expands up to 4 μm in our sample, and a coherence time in the same order of magnitude as the polariton lifetime. In the presence of an externally applied magnetic field, we demonstrate the lifting of the valley degeneracy of the condensed state, and the valley polarization of the system.

We believe, that the giant exciton binding energies provided by TMDC layers will allow the observation of spatially and temporally coherent valley condensates at ambient conditions in carefully adapted devices. In conjunction with recently developed technologies for large-scale monolayer growth and electrical injection[35,36], this will represent an entirely new platform for coherent light-sources based on bosonic stimulation. The capability to address and manipulate the valley index of the condensate makes our system highly promising for the exploration of large-scale coherent valley currents based on liquid light.

**Acknowledgement**

The authors gratefully acknowledge funding by the State of Bavaria and Lower Saxony. Funding provided by the European Research Council (ERC project 679288, unlimit-2D) is acknowledged. ST acknowledges funding from NSF DMR 1955889, DMR 1933214, and 1904716. ST also acknowledges DOE-SC0020653. ES and AVK acknowledge Westlake University (Project No. 041020100118) and the Program 2018R01002 funded by Leading Innovative and Entrepreneur Team Introduction Program of Zhejiang. K.W. and T.T.

acknowledge support from the Elemental Strategy Initiative conducted by the MEXT, Japan, Grant Number JPMXP0112101001, JSPS KAKENHI Grant Number JP20H00354 and the CREST(JPMJCR15F3), JST.

## Supplementary materials

### S1. Description of the coupled oscillator model

To describe the upper and lower polariton resonances, we employ a standard two-coupled-oscillators model:

$$\begin{bmatrix} E_{ex} & V/2 \\ V/2 & E_{cav} \end{bmatrix} \begin{bmatrix} X \\ C \end{bmatrix} = E \begin{bmatrix} X \\ C \end{bmatrix} \quad (S.1.1)$$

where $E_{ex}$ and $E_{cav}$ denote the energies of the exciton and cavity modes, respectively, and $V$ is the normal mode splitting. For the lower polariton branch the Hopfield coefficients $X$ and $C$ are given by:

$$|X|^2 = \frac{1}{2}\left(1 + \frac{E_{cav} - E_{exc}}{\sqrt{(E_{cav} - E_{exc})^2 + V^2}}\right) \quad (S.1.2)$$

$$|C|^2 = 1 - |X|^2 \quad (S.1.3)$$

Their squared amplitudes $|X|^2$ and $|C|^2$ quantify the exciton and cavity photon fractions. The eigenenergies of the upper and lower polariton branches are obtained by solving the eigenvalue problem:

$$E_{UP,LP}(k_{\parallel}) = \frac{1}{2}\left(E_{ex} + E_{cav} \pm \sqrt{V^2 + (E_{cav} - E_{exc})^2}\right) \quad (S.1.4)$$

### S2. Modelling magnetic effects on condensed TMD polaritons

For simulating the effect of the external magnetic field on the circular polarization degree of the polariton state, we use the model proposed in Ref. [34] in the main text. The model represents the system of coupled Boltzmann equations for the population of the left- and right-circularly polarized polaritons in the non-condensed and condensed states, $N_{R,C}^{\pm}(t)$:

$$d_t N_R^{\pm} = P_{\pm} - R N_R^{\pm}(N_C^{\pm} + 1) - \Gamma_R N_R^{\pm} \mp (\gamma_R^+ N_R^+ - \gamma_R^- N_R^-),$$

$$d_t N_C^{\pm} = R N_R^{\pm}(N_C^{\pm} + 1) - \Gamma_C N_C^{\pm} \mp (\gamma_C^+ N_C^+ - \gamma_C^- N_C^-),$$

The system is excited with the non-resonant optical pump $P_{\pm}$ which is taken to be linearly polarized, $P_+ = P_-$. The polariton condensate is fed from the non-condensed state with the rate $R$. $\Gamma_{C,R}$ are the decay rates. The effects of the external magnetic field and the polariton interactions are taken into account in the spin relaxation rates for polaritons in the condensed and non-condensed states as follows: $\gamma_C^{\pm} = \gamma_C \exp(\pm \Delta E_C / k_B T)$ and $\gamma_R^{\pm} = \gamma_R \exp(\pm \Delta E_R / k_B T)$, where $\gamma_{C,R}$ are the spin relaxation rates in the absence of the magnetic field, $\Delta E_C = E_C^+ - E_C^-$ and $\Delta E_R = E_R^+ - E_R^-$ are the splittings of energy of the spin-polarization components of polaritons in the condensed and non-condensed states, respectively. The energies are calculated as $E_C^{\pm} = E_{C0} \mp \frac{1}{2} X \beta B + \alpha X(N_R^{\pm} + X N_C^{\pm})$ and $E_R^{\pm} = E_{R0} \mp \frac{1}{2}\beta B + \alpha(N_R^+ + N_R^- + X N_C^{\pm})$, where $E_{C0}$ and $E_{R0}$ are energy levels at the zero magnetic field, $X$ is the exciton fraction in polaritons in the condensed state.

The constant $\beta$ is the induced Zeeman splitting per one Tesla used as a fitting coefficient. $\alpha$ is the interaction constant. In our simulations we take the following values of the parameters: the scattering rate is $R = 5 \times 10^{-3}$ ps$^{-1}$, the decay rates are $\Gamma_C = 6$ ps$^{-1}$ and $\Gamma_R = 2.5$ ps$^{-1}$, the spin relaxation rates are $\gamma_C = 0.1$ ps$^{-1}$ and $\gamma_R = 0.001$ ps$^{-1}$, the splitting constant is $\beta = 65$ μeV T$^{-1}$, the interaction constant is $\alpha = 10$ μeV, the exciton fraction is taken as $X = 0.5$.

## S3. Position-resolved cavity photoluminescence

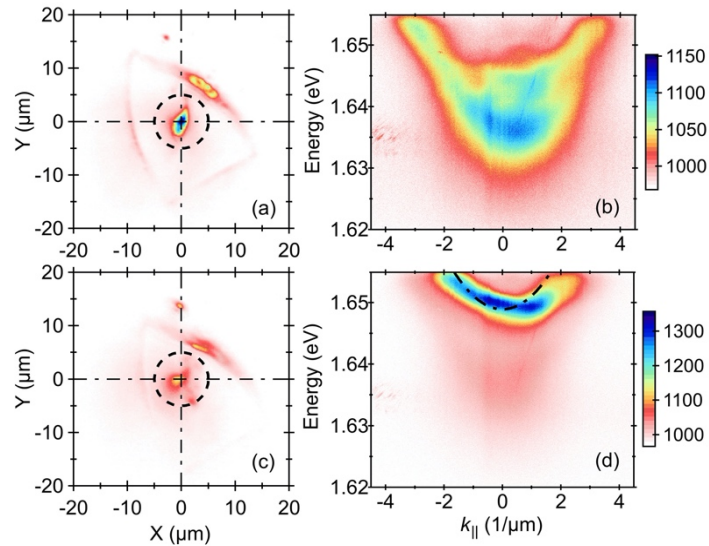

Fig S1. (a/c) Real space emission map of polaritons excited under the same conditions as those described in Fig. 4, with the pump spot located on/off the MoSe$_2$ monolayer. (b/d) Corresponding polariton/cavity dispersion relation on/off the flake. The dispersion relations have been filtered in real space, see dashed black circle indicating the real space area that has been resolved in momentum space. This comparison between these two dispersion relations allows to identify clearly the cavity mode and its curvature.

## S4. Photoluminescence excitation on different reference samples

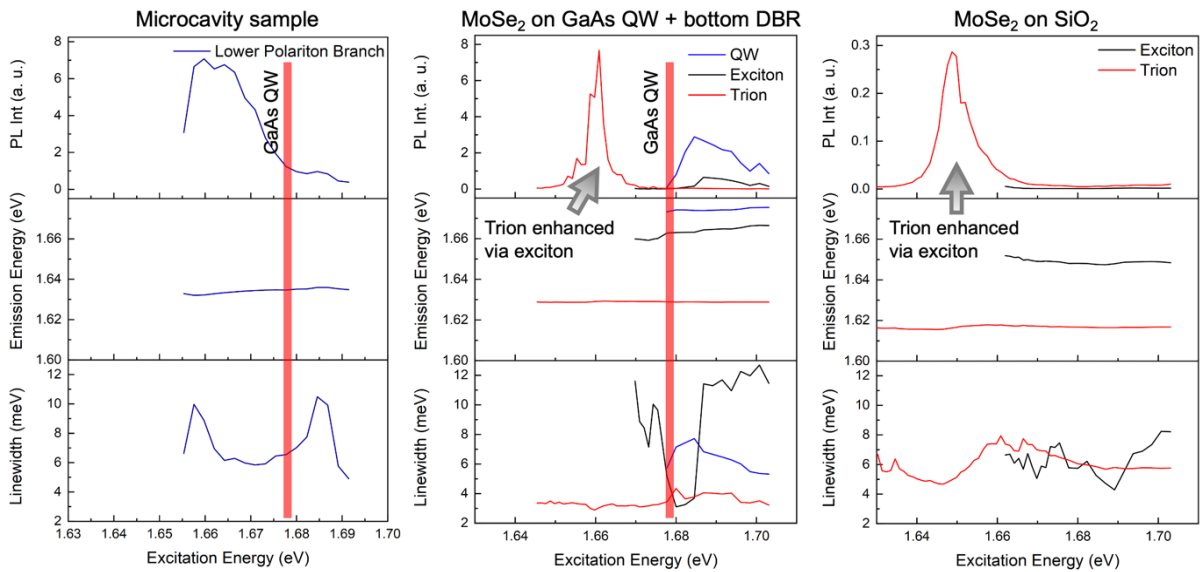

Fig S2. **Addressing the role of the GaAs quantum well in the polariton emission**. PLE study of three different samples, (left) MoSe$_2$ monolayer coupled to a microcavity, (center) MoSe$_2$ monolayer placed on top of the GaAs quantum well and bottom DBR, and (right) MoSe$_2$ monolayer deposited on SiO$_2$. Each column is subdivided in three panels containing (from top to bottom) the intensity, energy and linewidth of the emission as a function of the excitation energy. The spectral position of the quantum well is indicated with a vertical red line in the left and center columns. This representation evidences that the quantum well does not increase the MoSe$_2$ emission either from the exciton or from the trion state (see top panels in left and center columns). The PLE study in the center and left column reveals that the resonant

excitation of the MoSe$_2$ exciton enhances the emission from the trion (see the red peak in the top panels). Same excitation conditions as those described for Fig. 2(b) of the main text.

## S5. Polariton spectrum analysis above threshold

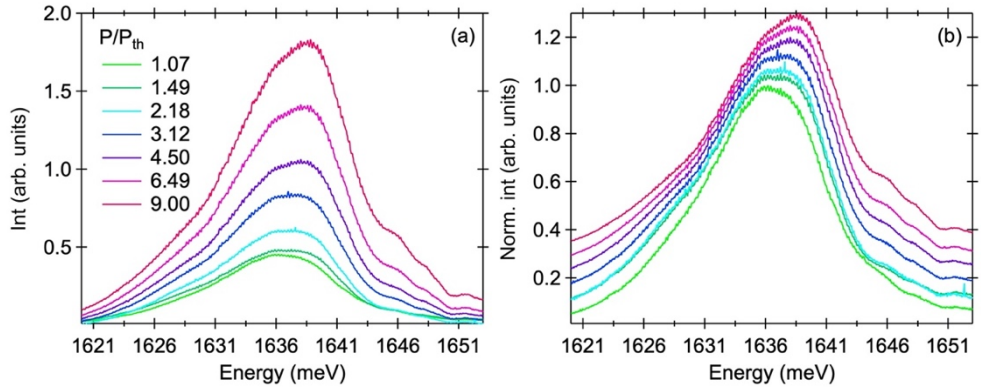

Fig. S3. Extracted from Fig. 3 of the main text, (a) polariton spectra at k$_∥$=0 for pump powers above threshold, (b) waterfall representation of the normalised spectra. From this analysis we observe the polariton emission blueshift above threshold: there is a competence between two main peaks, located at ~1636.3 and ~1638.3 meV. From panel b, we observe that the peak at 1636.3 meV remains at a constant energy, while the high energy peak blueshifts from 1637.3 to 1638.8 meV.

## S6.1 Dispersion relation and magnetic response below threshold

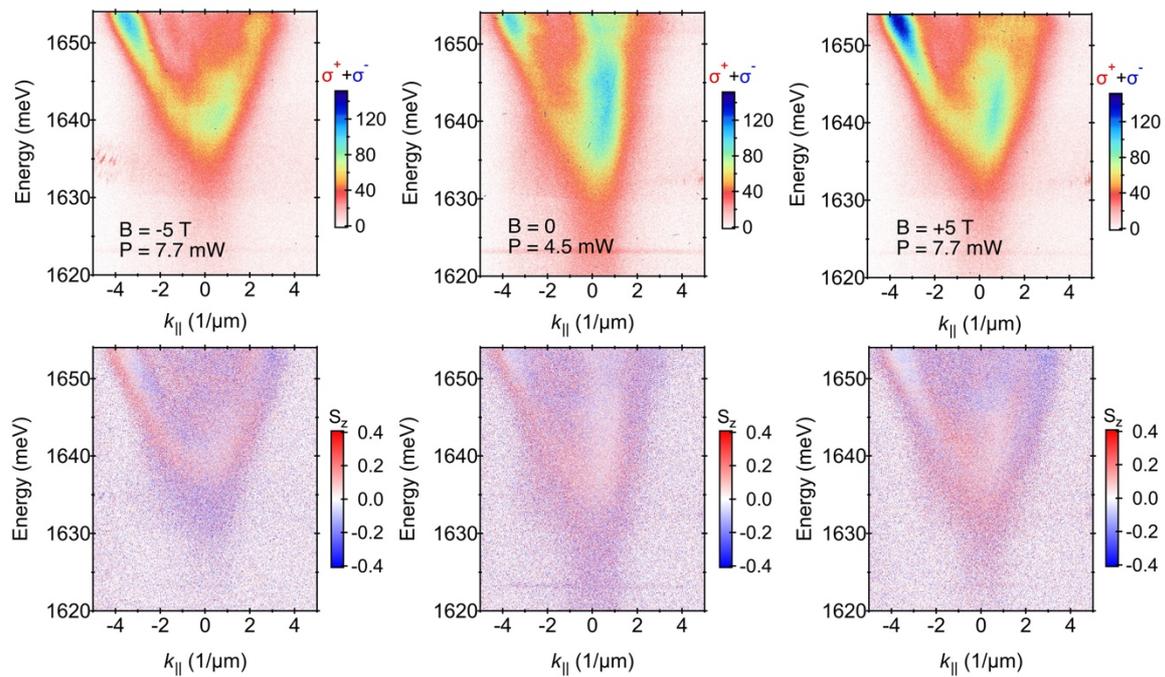

Fig. S4. Dispersion relation and circular degree of polarisation below threshold and as a function of magnetic field. (top row) Intensity map, σ$^+$+σ$^-$, and (bottom row) circular degree of polarization, S$_z$, as a function of energy and momentum (k$_∥$) of the polariton condensate (each pump power indicated in top panels), encoded in two different colour scales, respectively. Left, middle and right columns correspond to a magnetic field of -5, 0 and 5 T.

### S6.2 Polariton spectrum analysis versus magnetic field above threshold

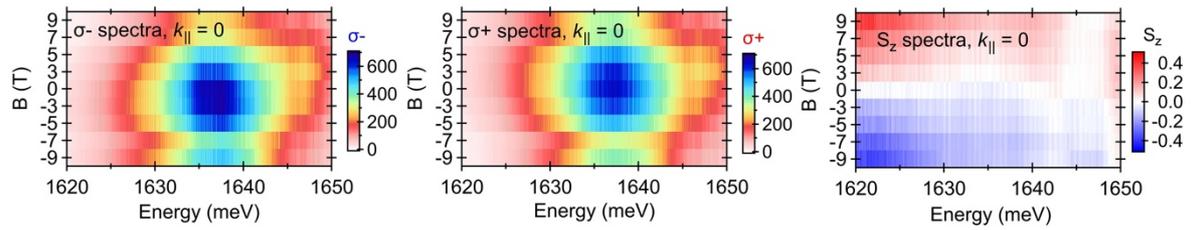

Fig. S5. (left, center panels) Intensity map of the σ⁺/σ⁻ spectrum recorded at $k_\parallel=0$ as a function of the magnetic field (vertical axis). (right panel) Degree of circular polarisation ($S_z$) at $k_\parallel=0$ as a function of energy and magnetic field, extracted from previous panels. Same excitation conditions as those described in Fig. 5 of the main text.